\documentclass{jpsj2}

\def\runtitle{Frequency Dependence of Quantum Localization}
\def\runauthor{Manabu \textsc{Machida}, 
Keiji \textsc{Saito}, and Seiji \textsc{Miyashita}}

\title{%
Frequency Dependence of Quantum Localization \\
in a Periodically Driven System
}

\author{%
Manabu \textsc{Machida}\thanks{machida@spin.t.u-tokyo.ac.jp}, 
Keiji \textsc{Saito}, and Seiji \textsc{Miyashita}
}

\inst{%
Department of Applied Physics, School of Engineering, 
The University of Tokyo, \\
7-3-1 Hongo, Bunkyo-ku, Tokyo 113-8656
}

\recdate{\today}

\abst{%
We study the quantum localization phenomena for a random matrix 
model belonging to the Gaussian orthogonal ensemble (GOE).  
An oscillating external field is applied on the system.
After the transient time evolution, energy is saturated 
to various values depending on the frequencies.
We investigate the frequency dependence of 
the saturated energy.  
This dependence cannot be explained by 
a naive picture of successive independent Landau-Zener transitions 
at avoided level crossing points.  
The effect of quantum interference is essential.
We define the number of Floquet states 
which have large overlap with the initial state, 
and calculate its frequency dependence.  
The number of Floquet states shows approximately linear dependence 
on the frequency, when the frequency is small.  
Comparing the localization length in Floquet states and that 
in energy states from the viewpoint of the Anderson localization, 
we conclude that the Landau-Zener picture works for the local 
transition processes between levels.  
}

\kword{%
quantum dynamics, quantum localization, random matrix, quantum chaos
}

\begin{document}
\sloppy
\maketitle

\section{Introduction}

The diffusive phenomena have been actively studied
in quantum systems with a complex level structure 
in the presence of a periodic external field.  
Under a periodic perturbation, when a system shows so-called 
quantum chaotic nature, it is widely believed that 
energy diffusion of the system eventually ceases 
and the energy saturates at a finite value.  
So far, only limited number of these systems have been actually studied.  
The kicked rotator model \cite{Casati79} and 
the kicked top model \cite{Haake87} 
are typical quantum models which show the quantum chaotic nature under 
a periodic external field.  The difference between classical and 
quantum energy diffusion has been discussed by investigating these 
models.  
In the quantum kicked rotator model, starting initially from 
the ground state, the system evolves diffusively 
only for a finite time.  After then the diffusive time-evolution 
ceases and the average energy is saturated to a finite value.  
That is, the system cannot absorb 
energy after that time.  This is a remarkable difference from 
the classical diffusion, where the energy increases unlimitedly.
The mechanism of this quantum localization in the model 
was discussed in the context of the Anderson localization
\cite{Fishman82,Grempel84}.  
Recently the kicked rotator model is experimentally 
realized using Hydrogen and Sodium atoms, and the 
quantum localization is observed
\cite{Galvez88,Bayfield89,Arndt91,Moore94,Collins95}.

It is known that random matrices well describe characteristics of 
the level statistics of the systems which show chaotic nature 
in the corresponding classical models\cite{Haake,Stoeckmann}.
Therefore we study the quantum localization 
process of random matrices to extract universal features.
In this paper, we investigate the time evolution of a system whose Hamiltonian 
is taken from the Gaussian orthogonal ensemble (GOE) 
with a periodic perturbation, and 
try to capture the universal features on the relation between 
the saturated energy and the frequency of the perturbation.
So far, several types of time-dependent random matrices 
are adopted to study the 
quantum diffusion process \cite{Cohen00,Wilkinson90,Bulgac96,Wilkinson88}.  
Wilkinson calculated the energy diffusion constant 
using the Landau-Zener transition 
formula\cite{Landau32,Zener32,Stueckelberg32} when the perturbation 
acts on the system linearly in time.\cite{Wilkinson88}  
He clarified the difference of diffusion constants between GOE and 
the Gaussian unitary ensemble (GUE).  Wilkinson neglected the effect 
of interference of quantum phase, 
which plays a crucial role in the quantum localization in the presence of
periodic perturbation \cite{Wilkinson90}. 
Cohen and Kottos considered the energy diffusion 
before the saturation in the 
Wigner's banded random matrix 
(WBRM) model with an oscillating perturbation, and calculated 
the diffusion coefficient as a function of the amplitude 
and frequency of the perturbation \cite{Cohen00}.  
Taking these previous studies into account, we study
how the Landau-Zener picture is related to the quantum 
localization by investigating the dependence of the saturated 
energy on the frequency of the perturbation. 

We first numerically demonstrate that the quantum localization occurs in 
the system, which cannot be explained by a naive application of 
an independent Landau-Zener picture to the system. 
In order to understand the quantum localization intuitively, we 
consider the overlap between Floquet eigenstates and the initial 
ground state, and find that the overlap exponentially decays in 
the Hilbert space spanned by Flouqet eigenstates, which directly 
means that the ground state is localized in this representation.  
We define the relevant number of Floquet eigenstates which has a large 
overlap with the initial ground state. This number was originally 
introduced as a quantity which corresponds to the Lyapunov exponent 
in the quantum kicked top model by Haake \textit{et al}.\cite{Haake87}.  
The dependence of the relevant number on the frequency is numerically 
investigated.  We numerically find that this number linearly depends 
on the frequency of the perturbation in small frequency regime.
Employing a plausible proportional relation between the localization 
length in the energy space and that in the Floquet space, which is 
also adopted in the case of the kicked rotator model, and 
taking account of this linear dependence, we find the Landau-Zener 
mechanism works in the view of the Anderson localization.  

This paper is organized in the following way.
In \S 2, we explain the random matrix model and the numerical method of time
evolution.  In \S 3, we study the frequency dependence of
the saturated average energy.  In \S 4, we discuss 
the quantum localization by introducing the number of relevant 
Floquet states.  Finally in \S 5, we give summary and 
discussion.

\section{Model and Method \label{mm}}

We consider time-reversal invariant systems under an oscillating 
external field.  The random matrix ensemble appropriate for describing 
the spectral statistics of these systems is the Gaussian orthogonal 
ensemble (GOE).  Random matrices well describe characteristics of energy 
spectra of complex quantum systems which have no conserved quantities. 

We shall consider the total Hamiltonian given by 
\begin{equation}
\mathcal{H}(t)=H_0+\lambda(t)V,
\label{mm02}
\end{equation}
where  $H_0$ denotes the non-perturbed part of the Hamiltonian, and  
$\lambda(t)V$ is the perturbation part with a time-dependent parameter 
$\lambda(t)$. We take $H_0$ and $V$ from the ensemble of the GOE random 
matrices with the dimension $N$.  
This model corresponds to the realistic systems such as 
billiard systems with a time-dependent boundary, or
complex spin systems with a time-dependent external field. 
Matrix elements ${H_0}_{ij}$ and $V_{ij}$ are taken from independent 
Gaussian random variables with mean 
zero and with the variance: 
${\rm var}\left({H_0}_{ij}\right) = 1 + \delta_{ij}$ and 
${\rm var}\left(V_{ij}\right) = 1 + \delta_{ij}$.  
Note that both of the non-perturbed and 
perturbed terms have the same statistical properties.  
The density of states $\rho (E)$ at the energy $E$ is given by Wigner's 
semicircle law for large $N$ \cite{Wigner59}, 
\begin{equation}
\rho(E)=\frac{1}{2\pi N}\sqrt{4N-E^2} .
\label{mm03}
\end{equation}

\begin{figure}[th]
\hspace*{8mm}
\begin{center}
\includegraphics[scale=0.5,clip]{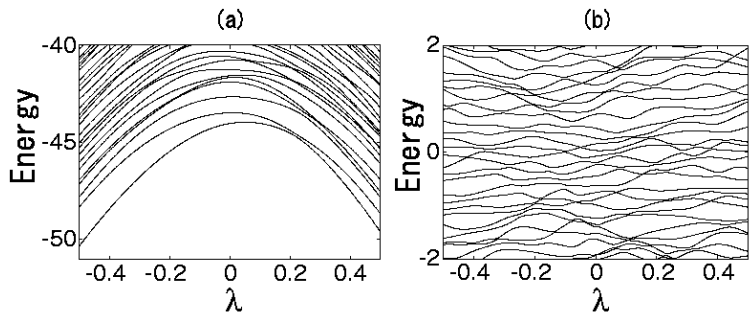}
\caption{Energy spectrum of $\mathcal{H}$ for $N=500$ 
as a function of $\lambda$ around 
(a) the ground state and (b) the center.}
\label{spectrum}
\end{center}
\vspace*{0mm}
\end{figure}

In this paper, we confine $\lambda(t)$ to the sinusoidal form, 
$ \lambda (t) = A\sin(\omega t)$ with $A=0.5$.  
Eigenvalues of a GOE random matrix show a structure with level
repulsion as a function of $\lambda$.  
Energy spectrum of a system of $N=500$ around the ground state, and 
around the center ($E=0$) are depicted in Fig.~\ref{spectrum} 
(a) and (b), respectively.  
Many avoided crossings are seen in Figs.~\ref{spectrum}.  There are 
no degeneracies of levels 
although some energy levels are seen to be crossing due to the line width.

Now we consider the time evolution of a state $|\psi\rangle$ 
of the system; 
\begin{equation}
{\rm i}\frac{\partial}{\partial t}|\psi\rangle = 
\mathcal{H}(t)|\psi\rangle.
\label{mm04}
\end{equation}
Here we take $\hbar=1$ and the initial 
state $|\psi_0\rangle$ is taken to be the ground state.  
The state $|\psi_1\rangle$ after a period $T=2\pi/\omega$ from 
$t=0$ is expressed using the Floquet operator\cite{Floquet83} $F$; 
\begin{equation}
|\psi_1\rangle = F|\psi_0\rangle,\quad 
F \equiv {\rm T}\exp\left[-{\rm i}\int_0^{2\pi/\omega}\mathcal{H}(t) 
{\rm d}t \right] ,
\label{fqt}
\end{equation}
where ${\rm T}$ means the time ordered product.  
Therefore the state after the $n$th period $|\psi_n\rangle$ is 
written as 
\begin{equation}
|\psi_n\rangle = F^n|\psi_0\rangle.
\label{mm05}
\end{equation}
Floquet eigenvectors $\left\{|\nu\rangle \right\}$ 
form a complete set in the 
Hilbert space, and Floquet eigenvalues are aligned on a
unit circle of the complex plane; 
\begin{equation}
F|\nu\rangle = e^{{\rm i}\phi_{\nu}}|\nu\rangle.
\label{mm06}
\end{equation}
We calculate $F$ numerically by integrating the Schr\"{o}dinger 
equation (\ref{mm04}) for a period using the fourth 
order decomposition of 
time-evolution operator \cite{Suzuki}.  
The time evolution of energy is expressed in terms of 
Floquet eigenvalues and eigenstates.  
The energy after the $n$th period is expressed
using the Floquet operator; 
\begin{align}
\langle\psi_n|\mathcal{H}|\psi_n\rangle &= 
\langle\psi_0|(F^{\dagger})^n\mathcal{H}F^n|\psi_0\rangle \nonumber \\
&= \sum_{\nu',\nu} e^{{\rm i}n(\phi_{\nu}-\phi_{\nu'})}
\langle\psi_0|\nu'\rangle  {}\langle\nu| \psi_0 \rangle
{}\langle\nu'|\mathcal{H}|\nu\rangle .
\label{mm07}
\end{align}

\section{$E_{\rm sat}$ as a Function of $\omega$ \label{esat}}
\subsection{Numerical results}

We define $g(\nu,\nu')$ and $E_{\rm sat}$ as, 
\begin{align}
g(\nu,\nu') &\equiv \langle\psi_0|\nu\rangle\langle\nu|H_0|\nu'\rangle
\langle\nu'|\psi_0\rangle , \nonumber \\
E_{\rm sat} &\equiv \sum_{\nu} g(\nu,\nu),
\label{esat01}
\end{align}
and rewrite eq.~(\ref{mm07}) as, 
\begin{align}
\langle\psi_n|H_0|\psi_n\rangle  
&= \sum_{\nu',\nu} e^{{\rm i }n(\phi_{\nu'}-\phi_{\nu})}
g(\nu,\nu') \nonumber \\
&= E_{\rm sat} + \sum_{\nu<\nu'}\left(
e^{{\rm i }n(\phi_{\nu'}-\phi_{\nu})}g(\nu,\nu')+\mbox{c.c.}\right).
\label{esat02}
\end{align}
In the second term of the right-hand side of eq.~(\ref{esat02}), 
the expectation values of terms for $\{\nu\ne\nu'\}$ oscillates 
with the period $2\pi/(\phi_{\nu'}-\phi_{\nu})$.  Thus, 
we see $\langle\psi_n|H_0|\psi_n\rangle$ fluctuates 
around $E_{\rm sat}$.  Therefore, 
$E_{\rm sat}$ is regarded as the saturated value of the energy. 

\begin{figure}[th]
\hspace*{8mm}
\begin{center}
\includegraphics[scale=0.5,clip]{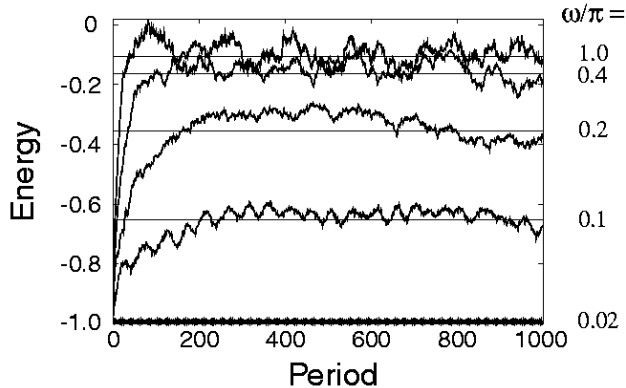}
\caption{Zigzag curves represent the time evolution of 
the average energy $E(t)$.   Horizontal lines represent 
$E_{\rm sat}$.  A pair of $E(t)$ and $E_{\rm sat}$ denotes, 
from the bottom, data for 
$\omega = $ $0.02\pi$, $0.1\pi$, $0.2\pi$, $0.4\pi$, and $1.0\pi$.}
\label{evol}
\end{center}
\vspace*{0mm}
\end{figure}

\begin{figure}[th]
\hspace*{8mm}
\begin{center}
\includegraphics[scale=0.5,clip]{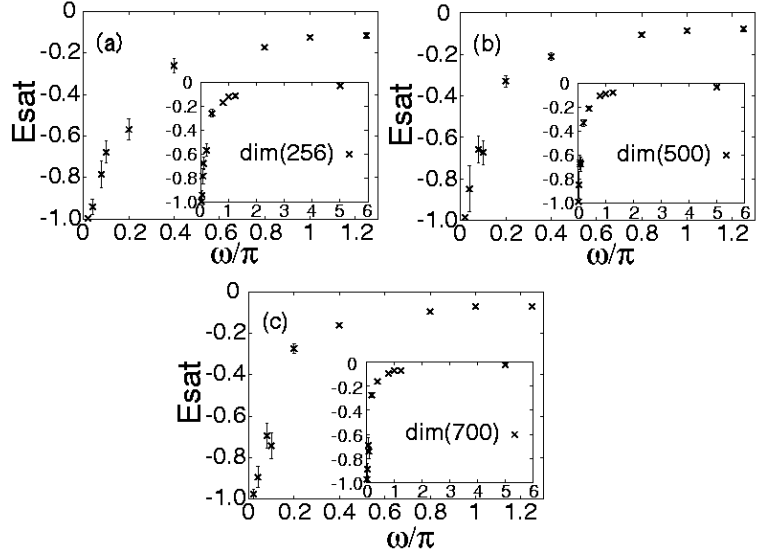}
\caption{$E_{\rm sat}$ as a function of $\omega$ for 
(a) $N=256$, (b) $N=500$, and (c) $N=700$.  
Errorbars show the variance among the data obtained from 
five samples.}
\label{nesat}
\end{center}
\vspace*{0mm}
\end{figure}

Figure \ref{evol} shows the time evolution of 
$\langle\psi_n|H_0|\psi_n\rangle$ (zigzag curve), 
and the corresponding $E_{\rm sat}$ 
(horizontal line)  with respect 
to $\omega/\pi=$ 0.02, 0.1, 0.2, 0.4, and 1.0 for $N=500$.  
Figure \ref{nesat} shows $E_{\rm sat}$ 
as a function of $\omega$ obtained from five samples for 
$N = 256$, $500$, and $700$.  
We should note that the maximal saturated energy is zero because 
the energy spectrum is symmetric about zero energy (E=0), 
and this maximal energy corresponds 
to the high temperature limit.  
Here we normalize $E_{\rm sat}$ such that 
the distance between $E=0$ and the ground state is $1$.  
Indeed in classical systems, the energy diffusion continues forever, 
which means that $E_{\rm sat}$ converges to $0$.  

On the other hand, when $\omega$ is very small ($\omega=0.02\pi$), 
the time evolution of the system is almost adiabatic, and the 
average energy changes little.\cite{Joye91}   When $\omega$ becomes
slightly larger ($\omega=0.1\pi \mbox{ and }0.2\pi$), 
transitions between levels begin to occur at 
avoided crossings (that is, nonadiabatic transitions).  
Thus the system absorbs energy and $E(t)$ increases.  However 
$E(t)$ saturates before it reaches to 0.
The value of $E_{\rm sat}$ gradually approaches to the center of the 
energy spectrum ($E_{\rm sat}=0$) as $\omega$ grows larger 
($\omega=0.4\pi \mbox{ and } 1.0\pi$).  
These finite saturations of $E_{\rm sat}$ are caused by the quantum 
effects, and this feature is called the quantum localization
\cite{Casati79,Haake}. The quantum localization was first discovered
in the kicked rotator model, which shows chaotic motion 
in the classical limit.\cite{Casati79}  


\subsection{Failure of a Picture of Independent Landau-Zener Transitions\label{lz}}

As shown in the previous section, the energy is saturated to a finite 
value.  We here show that this saturation cannot be realized 
without considering the quantum interference effect.  Let us consider 
the small frequency regime where transitions between energy levels 
take place only between the two levels at an avoided level 
crossing point.  
In such a region, we can introduce the well-known Landau-Zener picture 
for each transition.  
Wilkinson proposed a theory of the evolution of the energy 
in a random matrix model with a time-dependent
perturbation.\cite{Wilkinson88}  His theory assumes that 
transitions take place only at avoided crossings 
and the transition probability is determined by the Landau-Zener 
formula\cite{Zener32}.  This implies that the sweeping speed of 
the parameter is sufficiently slow, and there multiple level 
scatterings are not relevant.   
In the case that the field increases linearly in time, 
he found that this approximation of independent Landau-Zener 
scattering gives a good result.\cite{Wilkinson90}

If we apply a simple independent Landau-Zener picture to our case 
, we will find,
\begin{equation}
E_{\rm sat} = 0 .
\label{lz01}
\end{equation}
In what follows, we derive eq.~(\ref{lz01}).  
We assume that each 
transition occurs at an avoided level crossing point and the 
probability is given 
by the Landau-Zener formula.  Let us consider the transition rate $R$ 
which is the probability per unit time that the system makes a 
transition to one of the two neighboring states.  This quantity $R$ 
was already given for the case where the external field increases 
linearly in time \cite{Wilkinson88}.  
We apply the known expression to the present oscillating case.
The transition rate $R$ is determined by statistical distributions of 
the energy gaps and the asymptotic slopes at avoided crossing points.   
The average speed of the parameter $\lambda(t)$ is denoted by 
$\bar{|\dot{\lambda}|}$.  
For the present oscillating field, $\bar{|\dot{\lambda}|}$ is 
roughly estimated as $\bar{|\dot{\lambda}|}\simeq A\omega$.  
In addition, let $\rho$ denote the density of states 
and $\bar{\sigma}$ denote the typical difference between 
the asymptotic slopes of an avoided crossing. 
Under the conditions of slow speed, 
$\hbar\bar{\sigma}|\bar{\dot{\lambda}}|\rho^2\ll 1$, which corresponds to the 
situation that transitions of states only occur at avoided crossings to 
the adjacent levels, $R$ is expressed as,
\begin{equation}
R \propto \hbar^{1/2}\bar{\sigma}^{1/2}|\bar{\dot{\lambda}}|^{3/2}\rho^{2} 
\propto \hbar^{1/2}|\bar{\dot{\lambda}}|^{3/2}\rho^{3/2}.
\label{lz02}
\end{equation}
We consider the time evolution of the occupation probability for 
the eigenstate with energy $E$ at time $t$, $f(E,t)$.  
We can derive the diffusion equation for the 
occupation probability $f(E,t)$ of the state by extending Wilkinson's 
theory to the case that $\rho(E)$ is not constant; 
\begin{equation}
\frac{\partial \rho f}{\partial t} + 
|\bar{\dot{\lambda}}|\frac{\partial}{\partial E}
\left(\frac{dE}{d\lambda}\rho f\right) 
=\frac{\partial}{\partial E}
\left[D\rho\frac{\partial f}{\partial E}\right],
\label{lz03}
\end{equation}
where $D=R/{\rho^2}$.  Using the equation of continuity: 
\begin{equation}
\frac{\partial \rho}{\partial t} + 
|\bar{\dot{\lambda}}|\frac{\partial}{\partial E}
\left(\frac{dE}{d\lambda}\rho\right) = 0,
\label{lz04}
\end{equation}
eq.~(\ref{lz03}) is rewritten as,
\begin{equation}
\rho\frac{\partial f}{\partial t} =  
-|\bar{\dot{\lambda}}|\frac{dE}{d\lambda}\rho\frac{\partial f}{\partial E} +  
\frac{\partial}{\partial E}
\left[D\rho\frac{\partial f}{\partial E}\right].
\label{lz05}
\end{equation}

Let us consider the steady state $f(E) = f(E,\infty)$.
Then the left-hand side of eq.~(\ref{lz05}) gives zero.  Note that 
$\rho$ is given by the semicircle law (\ref{mm03}), 
$\rho(E)=\frac{1}{2\pi N}\sqrt{L^2-E^2}\,(L=2\sqrt{N})$ 
when $N$ is very large.  
Therefore, $f(E)$ is solved as 
\begin{equation}
f(E)=\int_{-L}^{E}c_1\exp\left[
\int_{-L}^{E'}k(\epsilon){\rm d}\epsilon\right]
{\rm d}E' + c_0,
\label{lz06}
\end{equation}
where $c_0$ and $c_1$ are constants of integration, and 
\begin{equation}
k(E) \equiv \frac{1}{D}\left(|\bar{\dot{\lambda}}|\frac{dE}{d\lambda}-
\rho^{-1}\frac{\partial(D\rho)}{\partial E}\right).
\label{lz07}
\end{equation}
When we integrate eq.~(\ref{lz06}), 
the second term of $k(\epsilon)$ diverges, because 
\begin{equation}
\int_{-L}^{E'}-\frac{1}{D\rho}\frac{\partial D\rho}{\partial\epsilon} 
{\rm d}\epsilon \propto 
-\int_{-L}^{E'}\frac{{\rm d}\ln\rho}{{\rm d}\epsilon} {\rm d}\epsilon
\rightarrow -\infty, 
\label{lz08}
\end{equation} 
where we used $D=R/\rho^2$ and eq.~(\ref{lz02}).  Therefore, 
\begin{equation}
f(E) = c_0.
\label{lz09}
\end{equation}
As a result, the average energy always becomes $0$ 
(the center of the energy spectrum); 
\begin{equation}
E_{\rm sat} = \int E \rho f(E) {\rm d}E = 0. 
\label{lz10}
\end{equation}
Thus we find that the idea of independent Landau-Zener scattering 
is not applicable at least to a periodic 
system.\cite{Wilkinson90,Gefen87}.  We cannot ignore the quantum 
mechanical interference effect for the quantum localization.

\subsection{Phenomenological Interpretation of the Quantum Localization 
in Analogy to the Anderson Localization \label{AL}}

We here phenomenologically interpret the quantum localization.
The quantum localization in the present system 
reminds us the Anderson localization where
a particle in the presence of a random potential is 
spatially localized \cite{Anderson58}.
The system for the Anderson localization is described by 
the Hamiltonian: 
\begin{equation}
\mathcal{H}_{\rm A} = \sum_i v_i |i\rangle \langle i| + 
t \sum_{\langle i,i'\rangle} |i\rangle \langle i'|.
\label{and01}
\end{equation}
Here, $|i\rangle$ means the state of the particle at the $i$th site, 
$v_i$ is the value of a random potential at the site $i$ 
which is uniformly distributed with a 
width $W$, and $t$ represents hopping to one of 
the nearest neighbor sites. In this particle system, the average 
hopping probability $p_{\rm A}$
is expressed as $p_{\rm A} \sim |t|/W$.  
The probability $p_{\rm A} (m)$ 
that the particle exists on the $m$th site from the 
initial ($0$th) site is estimated as $p_{\rm A} (m)\sim p_{\rm A}^{m}\sim
\exp\left(-m\log(W/|t|)\right)$.\cite{Nagaoka85}  
That is, the particle is localized in the space exponentially.
This localization is caused by the quantum interference effects.

\begin{figure}[th]
\hspace*{8mm}
\begin{center}
\includegraphics[scale=0.5,clip]{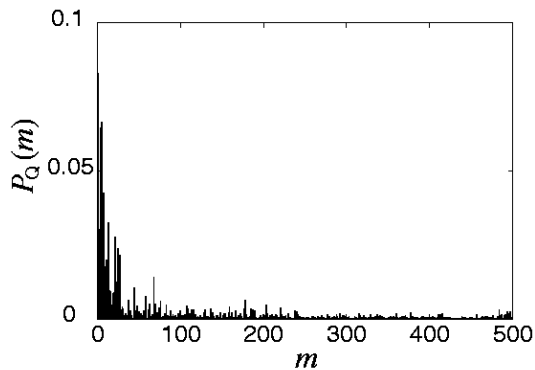}
\caption{The probability $P_{\rm Q}(m)$ that the state is 
on the $m$th level of the system (\ref{mm02}) for $N=500$ and 
$\omega=0.1\pi$.  It is obtained after 600 periods.} 
\label{anderson}
\end{center}
\vspace*{0mm}
\end{figure}

Let us consider the analogy to the Anderson localization in 
the quantum localization of the present model.  
We associate the state $|i\rangle$ in eq.~(\ref{and01}) 
with the $i$th adiabatic state of $H_{0}$.
Furthermore we focus on the frequency regime where 
transitions between levels take place only at avoided crossings.
We numerically confirmed that this situation is actually realized 
for small $\omega$. In this case, we define a 
transition probability $p_{\rm Q}$ between adjacent levels, 
which corresponds to $p_{\rm A}$ in the Anderson localization, and 
we find the common characteristics between the present 
situation in the random matrices and the Anderson localization.  
That is, the localization occurs
due to the quantum inference effect among many transitions between 
states.  We may write the occupation probability 
$p_{\rm Q}(m)$ that a state is on the $m$th level as 
$p_{\rm Q}(m) \sim p_{\rm Q}^{m}$, where $p_{\rm Q}$ is a transition 
probability between the states.  Actually fast relaxation is 
observed numerically in $p_{\rm Q}(m)$ of a typical example, which 
is depicted in Fig.~\ref{anderson}. 

If we assume an exponential form in the analogy to the Anderson 
localization, the saturation energy $E_{\rm sat}$ is written as,
\begin{equation}
E_{\rm sat} = \sum_{m=1}^{N} p_{\rm Q}^{(m-1)} E_{m}   /Z,
\label{and04}
\end{equation}
where $E_{m}$ is the $m$th eigenvalue of $H_{0}$ and 
$Z$ is the normalization factor, 
$Z=\sum_{m=1}^{N}p_{\rm Q}^{m-1}$.  
Equation (\ref{and04}) gives a finite saturation value of 
$E_{\rm sat}$.

\section{The Relevant Number of Floquet Eigenstates}

In previous sections, we showed the quantum localization 
in the random matrix model and considered an intuitive interpretation. 
In this section, we consider the quantum localization from 
the viewpoint of the Floquet theory.  
As seen in eq.~(\ref{esat01}), the quantum dynamics is determined 
by the overlaps between the initial state and Floquet eigenstates, 
\textit{i.e.}, $|\langle \nu | \psi_{0} \rangle |^2$, and 
$E_{\rm sat}$ is obtained by Floquet eigenstates.   
Thus we here discuss this overlap.
\begin{figure}[th]
\hspace*{8mm}
\begin{center}
\includegraphics[scale=0.5,clip]{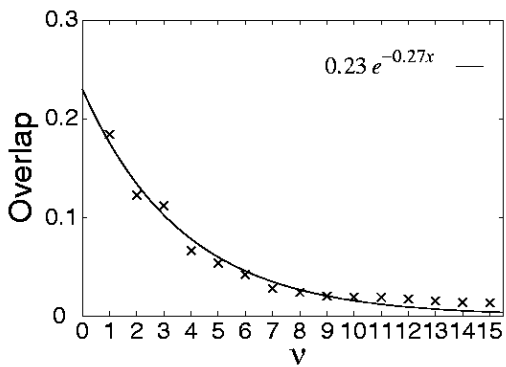}
\caption{The dependence of the overlap $|\langle\nu|\psi_0\rangle|^2$ 
on the number $\nu$.  $|\langle\nu|\psi_0\rangle|^2$ is arranged 
 in order of the magnitude.  The overlap decays exponentially.  
Here $\omega=0.1\pi$ and $N=500$.}
\label{overlap}
\end{center}
\vspace*{0mm}
\end{figure}

\begin{figure}[th]
\hspace*{8mm}
\begin{center}
\includegraphics[scale=0.5,clip]{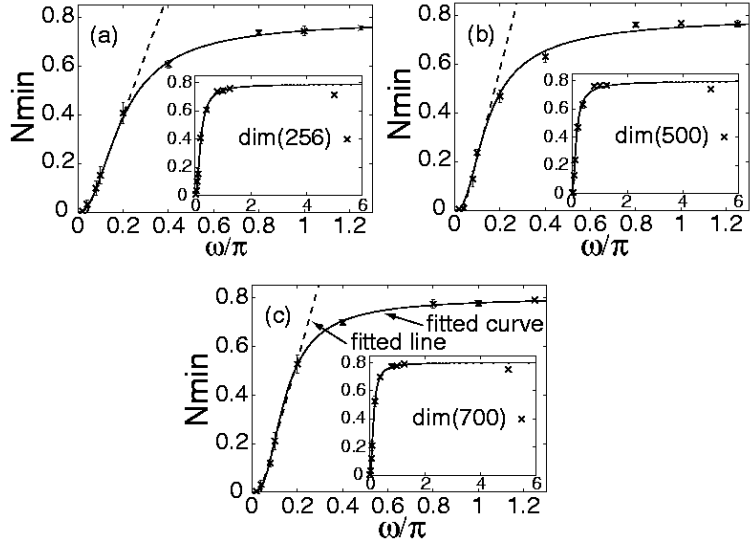}
\caption{$N_{\rm min}$ as a function of $\omega/\pi$ for 
(a) $N=256$, (b) $N=500$, and (c) $N=700$.  
Errorbars show the variance among the data obtained from five 
samples.  The fitted line and also linear fitting in the intermediate 
region are shown in each graph.}
\label{nminw}
\end{center}
\vspace*{0mm}
\end{figure}

We found numerically that the distribution of 
$|\langle\nu|\psi_0\rangle|^2$ decays approximately following an 
exponential function when we arrange the states 
in order of the magnitude of the overlap as is shown in 
Fig.~\ref{overlap}.  This property is one of the characteristics 
of the quantum localization.  
We study how many Floquet states are involved in the ground state, and 
define the minimal number $N_{\rm min}$ of the Floquet states 
by which the initial state $|\psi_0\rangle$ is covered within a 
ratio $r$:
\begin{equation}
N_{\rm min} = 
\frac{1}{N}\min\left\{ \mathcal{N}_{\rm min}: 
\sum_{\nu=1}^{\mathcal{N}_{\rm min}}
|\langle\psi_0|\nu\rangle|^2>r; 
|\langle\psi_0|\nu\rangle|^2\ge |\langle\psi_0|\nu+1\rangle|^2\right\}.
\label{nmin01}
\end{equation}
This quantity is the same as the quantity used in the 
different context by Haake, Kus, and Scharf.\cite{Haake87}  
Here, we take $r=0.99$.  
Figure \ref{nminw} shows $N_{\rm min}$ as a function of 
$\omega/\pi$ obtained from five samples for $N=256$, $500$, and $700$.  

We introduce the fitting function for $N_{\rm min}(\omega)$ 
in the form: 
\begin{equation}
N_{\rm min}(\omega) = N^*_{\rm min}\left\{1-\frac{1}{\left[
(\omega/\omega^*)^a+1\right]^b}\right\}.
\label{nmin02}
\end{equation}
This function fits the data quite well.
In eq.~(\ref{nmin02}), $N^*_{\rm min}$, $a$, $b$, and $\omega^*$ 
are determined from the least-squares method.  
The results are shown in Table \ref{fit_Nmin}.  
From eq.~(\ref{nmin02}), $N_{\rm min}(\omega)$ is extremely small 
when $\omega \ll \omega^{*}$.  
This is partially because the energy gaps in lower 
energy region are so large comparing with $\omega$ that 
transition to higher levels from the ground state seldom 
occurs (see Fig.~\ref{spectrum}(a)), 
and therefore the system behaves almost adiabatically.\cite{Joye91}  
On the other hand, $N_{\rm min}(\omega)$ is 
well fitted by a linear function (dashed line) for the region of 
$\omega \sim \omega^{*}$.

\begin{table}[th]
\caption{Values of the parameters in eq.~(\ref{nmin02}) 
obtained by the least-squares method.}
\label{fit_Nmin}
\begin{halftabular}{@{\hspace{\tabcolsep}\extracolsep{\fill}}cccc} \hline
variables&dim(256)&dim(500)&dim(700) \\ \hline
$N^*_{\rm min}$ & 0.79     & 0.80     & 0.80     \\ \hline
a               & 2.3      & 2.6      & 3.0      \\ \hline
b               & 0.67     & 0.47     & 0.54     \\ \hline
$\omega^*/\pi$  & 0.15     & 0.10     & 0.11     \\ \hline
\end{halftabular}
\end{table}

So far, we observed localization in the energy space and also in the 
space of Floquet states.  Now we discuss on localization length 
of the two localizations.  In particular, we focus on the linear 
dependence of $N_{\rm min}(\omega)$ on $\omega$ in the small $\omega$ 
region.  We employ a phenomenological argument using the analogy to the 
Anderson localization as explained in \S\ref{AL}.  
Let us express the transition probability $p_{\rm Q}$ using 
some function $\Gamma(\omega)$ in the form: 
\begin{equation}
p_{\rm Q} \sim\exp\left[-\Gamma^{-1}(\omega)\right].
\label{nmin03}
\end{equation}
The occupation probability $p_{\rm Q}(m)$ that a stationary state 
is on the $m$th level is expressed as $
p_{\rm Q}(m) \sim p_{\rm Q}^{m} \sim \exp\left[-\Gamma^{-1}(\omega)m\right]$.  
Thus we can regard $\Gamma(\omega)$ as the localization length 
in the energy space.  
Now let $l$ denote the localization length in the Floquet space, 
\textit{i.e.}, $|\langle\nu|\psi_0\rangle|^2 \simeq c e^{-\nu/l}$, where
$c$ is the normalization constant. From the definition (\ref{nmin01}), 
$N_{\rm min}$ is expressed using the length $l$ as
\begin{equation}
N_{\rm min} \simeq -\frac{l}{N}\ln(1-r) \propto l.
\label{nmin04}
\end{equation}
Since we now focus on the linear dependence of $N_{\rm min}$ on 
$\omega$, we write, 
\begin{equation}
N_{\rm min} = \alpha \omega + \beta ,
\label{nmin05}
\end{equation}
where $\alpha$ and $\beta$ are constants.  
We note that $\beta$ is negligibly small.  Then we have 
\begin{equation}
l \propto \omega.
\label{nmin06}
\end{equation}
As has been assumed in the the kicked rotator model 
to discuss the diffusion constant \cite{Graham90}, 
we assume the proportionality between $l$ and $\Gamma(\omega)$, 
\begin{equation}
l \propto \Gamma(\omega).
\label{nmin07}
\end{equation}
Then, we obtain from eqs.~(\ref{nmin06}) and (\ref{nmin07}), 
\begin{equation}
\Gamma(\omega) = \omega/h.
\label{nmin08}
\end{equation}
Here $h$ is an unknown amplitude which may depend on $N$, $A$, 
the typical size $\bar{\epsilon}$ of gaps at avoided crossings, 
and the difference $\bar{\sigma}$ of the two asymptotic slopes at an 
avoided crossing.  
This $\omega$ dependence is consistent with that of the Landau-Zener 
formula, because eqs.~(\ref{nmin03}) and (\ref{nmin08}) give 
\begin{equation}
p_{\rm Q} \sim \exp\left[-\frac{h}{\omega}\right].
\label{nmin09}
\end{equation}
Actually, we numerically confirmed that transitions between 
levels take place only at avoided crossings in the region where 
$N_{\rm min}$ linearly depends on $\omega$.  
Thus we suppose that the amplitude of local transition probability is 
originated in the Landau-Zener transition, although the phase 
interference has an important effect on the global phenomena.

\section{Summary and Discussion}

We have studied numerically the time evolution of the average 
energy of the GOE random matrix model under a perturbation of a 
sinusoidal function of time.  In \S 3, we showed that the quantum 
localization occurs in this model.  We have also studied the frequency 
dependence of the saturated energy $E_{\rm sat}$, and we showed that 
we can not rely on the picture of independent occurrence of 
Landau-Zener transitions in the present case because the suppression 
of the quantum diffusion is essentially due to the phase interference 
effect of quantum mechanics.  We discuss this quantum localization in 
analogy to the Anderson localization.  
In \S 4, we introduced the relevant number of Floquet states 
$N_{\rm min}$, and proposed a form of $N_{\rm min}$ as a function 
of $\omega$.  $N_{\rm min}$ is almost zero when $\omega$ is nearly zero, 
and it saturates when $\omega$ is sufficiently large.  In the 
intermediate region, $N_{\rm min}$ grows linearly as $\omega$ increases.  
This linear dependence of $N_{\rm min}$ on $\omega$ implies that the 
localization length $l$ in the Floquet eigenstates also depends 
linearly on $\omega$ (eq.~(\ref{nmin04})).  On condition that 
$\Gamma\propto l$, the localization length $\Gamma$ of the eigenstates 
of the Hamiltonian is also proportional to $\omega$.  This dependence 
of $\Gamma$ on $\omega$ implies the Landau-Zener mechanism governs 
the local transitions.

\section*{Acknowledgments}
This study is partially supported by Grant-in-Aid from the Ministry 
of Education, Culture, Sports, Science and Technology of Japan.  
The computer calculation was partially carried out at the computer
center of the ISSP, which is gratefully acknowledged.

\end{document}